\documentclass{article}[12pt]
\usepackage{amsmath,amssymb}
\usepackage{graphicx}%
\usepackage{verbatim}

\begin{document}

\begin{center}
{\bf \Large Qualitative and Numerical Analysis of \\[6pt]
a Cosmological Model Based on an Asymmetric\\[6pt]
Scalar Doublet with Minimal connections.\\[6pt]
IV. Numerical Modeling and Types of Behavior of the Model                                                                                    } \\[12pt]
Yu. G. Ignat'ev and I. A. Kokh\\
N. I. Lobachevsky Institute of Mathematics and Mechanics of Kazan Federal University,\\
Kremleovskaya str., 35, Kazan, 420008, Russia.
\end{center}


\begin{abstract}

On the basis of a qualitative and numerical analysis of a cosmological model based on an asymmetric scalar doublet of nonlinear, minimally interacting scalar fields -- one classical and one phantom, peculiarities of the behavior of the model near zero energy hypersurfaces have been revealed.  Numerical models have been constructed, in which the dynamical system has limit cycles on the zero-energy hypersurfaces. Three types of behavior of the cosmological model have been distinguished, configured by the fundamental constants of the scalar fields and the initial conditions.  It is shown that over a wide sector of values of the fundamental constants and initial conditions, the cosmological models have a tendency to adhere to the zero-energy hypersurfaces corresponding to  4-dimensional Euclidean space. \\


\end{abstract}

\section{Introduction}

\noindent In our previous papers [1--3] we presented a qualitative analysis of a cosmological model, based on an asymmetric scalar doublet, and an analytical and numerical study of its behavior near zero effective energy hypersurfaces.  In [3] we also introduced maps of singular points of a dynamical system described by a normal autonomous system of ordinary differential equations [1]: 
\begin{equation} \label{_1_} 
\begin{array}{c} {\Phi '=Z,} \\[12pt] {Z'=-\sqrt{3} Z\sqrt{\left(Z^{2} +e\Phi ^{2} -{\displaystyle\frac{\alpha _{m} }{2}} \Phi ^{4} \right)-\left(z^{2} -\varepsilon \mu ^{2} \varphi ^{2} +{\displaystyle\frac{\beta _{m} }{2}} \varphi ^{4} \right)+\lambda _{m} } -e\Phi +\alpha _{m} \Phi ^{3} ,} \\[12pt] {\varphi '=z,} \\[12pt] {z'=-\sqrt{3} z\sqrt{\left(Z^{2} +e\Phi ^{2} -{\displaystyle\frac{\alpha _{m} }{2}} \Phi ^{4} \right)-\left(z^{2} -\varepsilon \mu ^{2} \varphi ^{2} +{\displaystyle\frac{\beta _{m} }{2}} \varphi ^{4} \right)+\lambda _{m} } +\varepsilon \mu ^{2} \varphi -\beta _{m} \varphi ^{3} ,} \end{array} 
\end{equation} 
where 
\[\alpha _{m} =\frac{\alpha }{m^{2} } ,\quad \beta _{m} =\frac{\beta }{m^{2} } ,\quad \lambda _{m} =\frac{\lambda }{m^{2} } ,\, \, \, \, \, \Lambda _{m} =\lambda _{m} -\frac{1}{2\alpha _{m} } -\frac{\mu ^{2} }{2\beta _{m} } ,\quad {\rm and}\quad \mu \equiv \frac{{\rm {\tt m}}}{m} \] 
are normalized, dimensionless parameters of the model, which we will assign in the form of a list ${\bf P}$
\begin{equation} \label{_2_} 
{\bf P}\equiv [\alpha _{m} ,\beta _{m} ,e,\varepsilon ,\mu ,\lambda _{m} ],\quad {\bf I}\equiv [\Phi (0),Z(0),\varphi (0),z(0)],    
\end{equation} 
with ${\bf I}$ being a list of initial conditions\footnote{\ Since\ system\ (1)\ is\ invariant\ with\ respect\ to\ shifts\ of\ dimensionless\ time\ $\tau$ ,\ the\ choice\ of\ the\ initial\ time\ $ \tau_0=0$\ has\ no\ meaning.\ \ }.

\noindent In the present paper, we carry out a detailed numerical study of the cosmological evolution of an asymmetric scalar doublet as a function of the parameters of the model, taking into account the results of the previous qualitative analysis. Note that the Hubble constant $H(t)$ and the invariant cosmological acceleration $\Omega $ are assigned by the formulas [3]\footnote{\ Note\ that\ the\ condition\ $H\geq 0  $,\ employed\ in\ our\ work,\ as\ shown\ in\ [4],\ can\ be\ lifted,\ at\ least\ for\ the\ case\ of\ an\ isolated\ classical\ field\ with\ a\ nonminimal\ interaction.\ }
\begin{equation} \label{_3_} 
H(t)=\frac{\dot{a}}{a} \ge 0,\, \, \; \; \Omega (t)=\frac{a\ddot{a}}{\dot{a}^{2} } \equiv 1+\frac{\dot{H}}{H^{2} } =-\frac{1}{2} (1+3{\rm \varkappa }),     
\end{equation} 
where ${\rm \varkappa }=p_{m} /{\rm {\mathcal E}}_{m} $ is the ratio of the effective pressure to the effective energy density -- it is the effective barotropic coefficient, and the effective energy density and pressure of the dynamical system have the form  
\[\begin{array}{c} {{\rm {\mathcal E}}_{m} (\Phi ,Z,\varphi ,z)\equiv \left(Z^{2} +e\Phi ^{2} -{\displaystyle\frac{\alpha _{m} }{2}} \Phi ^{4} \right)+\left(-z^{2} +\varepsilon \mu ^{2} \varphi ^{2} -{\displaystyle\frac{\beta _{m} }{2}} \varphi ^{4} \right)+\lambda _{m} ,} \\ {{\rm }p_{m} (\Phi ,Z,\varphi ,z)\equiv \left(Z^{2} -e\Phi ^{2} +{\displaystyle\frac{\alpha _{m} }{2}} \Phi ^{4} \right)-\left(z^{2} +\varepsilon \mu ^{2} \varphi ^{2} -\displaystyle\frac{\beta _{m} }{2} \varphi ^{4} \right)-\lambda _{m} .} \end{array}\] 

\section{Numerical modeling of the dynamical system: $\lambda =0$ }

\noindent We present results of a numerical integration demonstrating the indicated peculiarities.  We note at once that the large number of parameters of the cosmological model based on an asymmetric doublet --  there are six of them in all -- makes sorting through all of the possible variants an extremely cumbersome task; therefore, we present only a few results below -- those which we think are the most interesting.  

\subsection{Case of accessibility of all singular points }
\begin{equation} \label{_4_} 
{\bf P}=[1,1,1,1,1,0].  
\end{equation} 

\subsubsection{General properties of the phase space}

\noindent In this case, the singular points have the following coordinates: 
\[M_{0} (0,0,0,0);\, \, \, \, \, M_{01} (1,0,0,0);\, \, \, \, \, M_{02} (-1,0,0,0),\] 
\begin{equation} \label{_5_} 
M_{10} (0,0,1,0);\, \, \, \, \, M_{20} (0,0,-1,0);\, \, \, \, \, M_{11} (1,0,1,0),   
\end{equation} 
\[M_{12} (1,0,-1,0);\, \, \, \, \, M_{21} (-1,0,1,0);\, \, \, \, \, M_{22} (-1,0,-1,0),\] 
and the invariant characteristics $\sigma _{i}^{2} $ are equal to [3]    
\[\sigma _{1}^{2} =\frac{3}{8} ,\quad \sigma _{2}^{2} =\frac{3}{8} ,\quad \sigma _{3}^{2} =\frac{3}{4} .\] 
All nine singular points of the dynamical system are accessible. The character of these points was represented in the scheme presented in Fig. 1 of [3].  The dependence of the boundaries of the forbidden regions of phase space in the projections $\Sigma _{\Phi } $ and $\Sigma _{\varphi } $ on the values of the dual potentials is shown in Fig. 1.   
 \begin{figure}[h!]
 \centerline{\includegraphics[width=15cm,height=5cm]{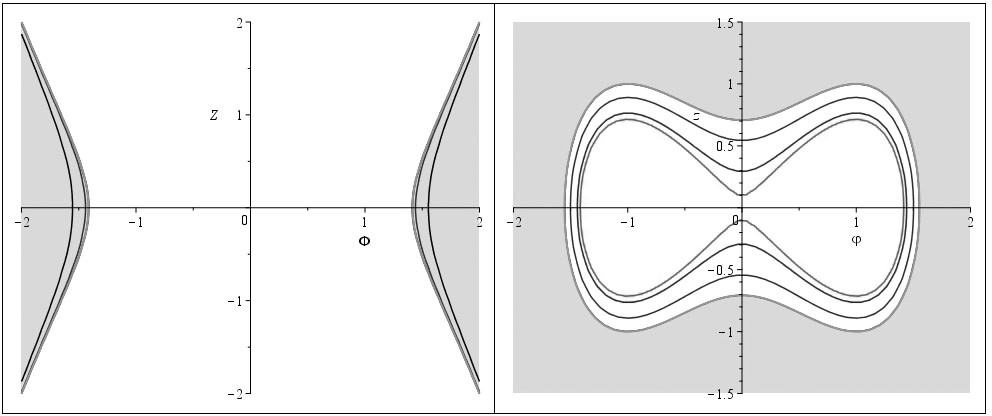}\label{Fig1}} \caption{  Dependence of the forbidden region (indicated by the light-gray color) in the projections $\Sigma _{\Phi } $ (left) and $\Sigma _{\varphi } $ (right) on the values of the dual potentials for the model parameters listed in Eq. \eqref{_4_}: left -- from the interior curves to the exterior curves: $\varphi =0.01,\, \, 0.1,\, \, 0.3\, \, {\rm and}\, \, 1$, the interior regions are forbidden, right -- from the interior curves to the exterior curves: $\Phi =0.1,\, \, 0.3,\, \, 0.6\, \, {\rm and}\, \, 1$, the exterior regions are forbidden. }
\end{figure}

\noindent Although the case under consideration is quite standard and does not contain any interesting features in the behavior of the model, we shall consider it in more detail in order to demonstrate general properties of the asymmetric scalar doublet model. 

\subsubsection{Phase trajectories of the dynamical system     }

\noindent Figures 2--5 present the results of numerical modeling of dynamical system \eqref{_1_} for the model parameters listed in Eq. \eqref{_4_} and the initial conditions   
\begin{equation} \label{_6_} 
{\bf I}=[0.7,0.5,0.01,0].   
\end{equation} 
Here and below, the circles correspond to the beginning and end of a phase trajectory, and the x's mark the singular points.  It can be seen in Fig. 2 that the phase trajectories in the $\{ \Phi ,Z\} $ plane rebound from the saddle points $M_{10} \; {\rm and}\; M_{20} $ and then spiral around the attractive focus $M_{0} $.  Simultaneously with this, the phase trajectories in the $\{ \varphi ,z\} $ plane rebound from the saddle point $M_{0} $ and then spiral around the attractive foci $M_{01} \; {\rm and}\; M_{02} $ (Fig. 3). This situation becomes clearer in the phase diagram in the plane of singular points $\{ \Phi ,\varphi \} $ (Fig. 4). Thus, the numerical results confirm the conclusions of the qualitative theory presented in the map of singular points in Fig. 1 of [3].  
 \begin{figure}[h!]
 \centerline{\includegraphics[width=15cm,height=5cm]{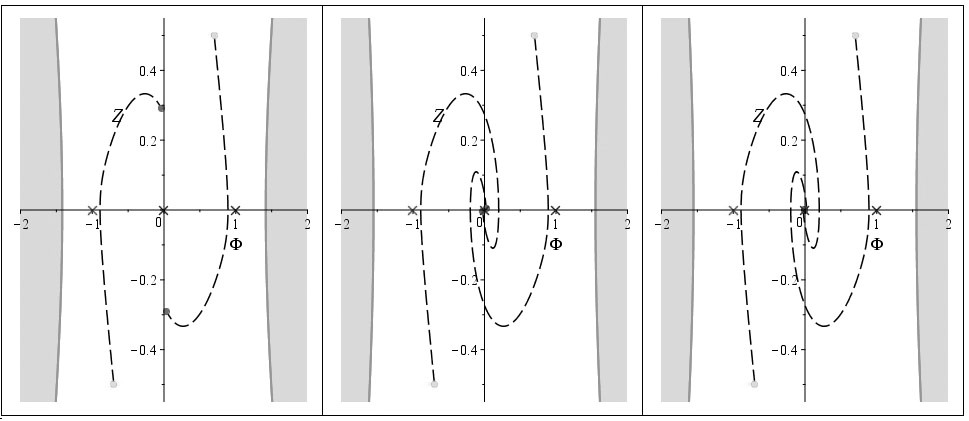}\label{Fig2}} \caption{  Cosmological evolution of a scalar doublet with the parameters listed in Eq. \eqref{_4_} and the initial conditions listed in Eq. \eqref{_6_} in the \textit{classical} plane $\Sigma _{\Phi } \equiv \{ \Phi ,Z\} $. The phase diagrams correspond to the times (from left to right) $\tau =5,\, \, 10\, \, {\rm and}\, \, 20$.  The forbidden regions correspond to the final time.}
\end{figure}
  \begin{figure}[h!]
 \centerline{\includegraphics[width=15cm,height=5cm]{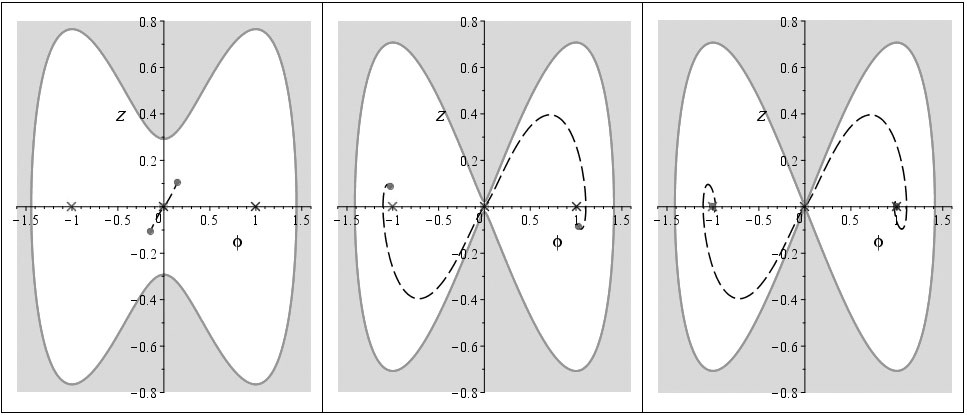}\label{Fig3}} \caption{ Cosmological evolution of a scalar doublet with the parameters listed in Eq. \eqref{_4_} and the initial conditions listed in Eq. \eqref{_6_} in the \textit{phantom} plane $\Sigma _{\varphi } \equiv \{ \varphi ,z\} $. The phase diagrams correspond to the times (from left to right) $\tau =5,\, \, 10\, \, {\rm and}\, \, 20$.  The forbidden regions correspond to the final time. }
\end{figure}
 \begin{figure}[h!]
 \centerline{\includegraphics[width=15cm,height=5cm]{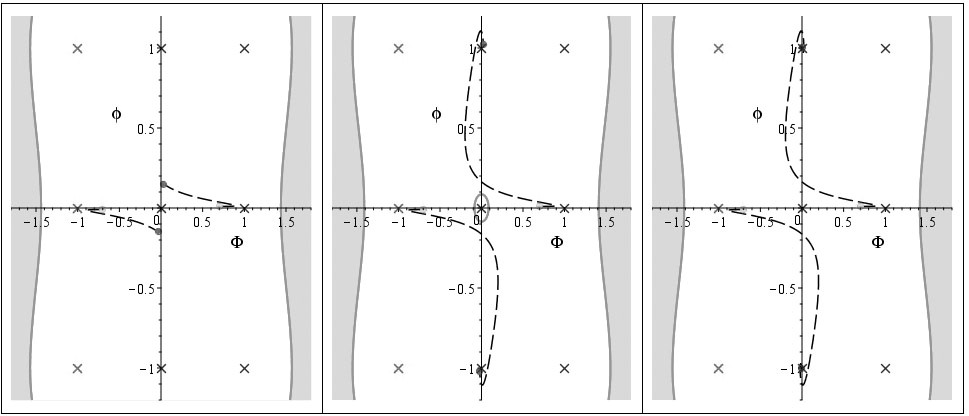}\label{Fig4}} \caption{ Cosmological evolution of a scalar doublet with the parameters listed in Eq. \eqref{_4_} and the initial conditions listed in Eq. \eqref{_6_} in the plane of potentials $\{ \Phi ,\varphi \} $. The phase diagrams correspond to the times (from left to right)  $\tau =5,\, \, 10\, \, {\rm and}\, \, 20$.}
\end{figure}
 \begin{figure}[h!]
 \centerline{\includegraphics[width=15cm,height=5cm]{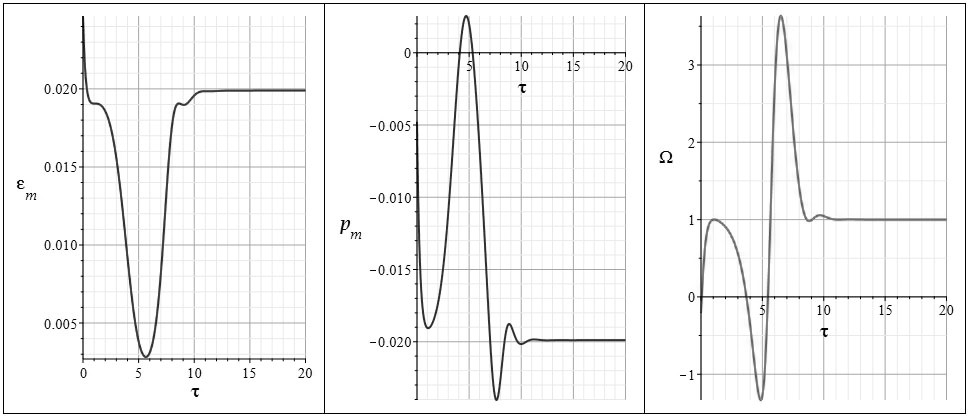}\label{Fig5}} \caption{ Cosmological evolution of the physical characteristics of the cosmological model with the parameters listed in Eq. \eqref{_4_} and the initial conditions listed in Eq. \eqref{_6_}.  From left to right: the dimensionless effective energy ${\mathcal E}_m$, the dimensionless effective pressure $p_{m} $, and the invariant cosmological acceleration $\Omega $.}
\end{figure}

 Figure 5 depicts the evolution of the physical characteristics of the cosmological model with the parameters listed in Eq. \eqref{_4_}.  Note that, according to the generally accepted classification, ${\rm \varkappa }$ takes the value ${\rm \varkappa }=0$ for nonrelativistic matter (in this case $\Omega =-1/2$), ${\rm \varkappa }=1/3$ for ultrarelativistic matter (in this case $\Omega =-1$), ${\rm \varkappa }=-1/3$ for quintessence (in this case $\Omega =0$), ${\rm \varkappa }=-1$ for pure inflation (in this case $\Omega =+1$), and ${\rm \varkappa }<-1$ for dark energy (in this case $\Omega >+1$).  As can be seen from Fig. 5, all of these values in the case under consideration after a burst tend toward constant values, and $\mathcal{E}_{m} +p_{m} \to 0$ as $\tau \to 0$, which corresponds to inflation in the final stage of evolution. Thus, the case under consideration is very close to the standard scenario with only one difference, namely that late inflation is supported by a phantom field, not a classical one.

The scenario considered above corresponds to small initial values of the phantom potential $\varphi (0)=\pm 0.01$.  An increase in the initial value of the phantom potential can substantially alter the cosmological scenario.  Let us specify the initial position of the dynamical system above the singular point $M_{01} $:
\begin{equation} \label{_7_} 
{\bf I}=[0.9,0.5,1.3,0.5]. 
\end{equation} 
In this case, we obtain phase diagrams that are fundamentally different from those considered above (Figs. 6--8):  the phase trajectories circumvent singular points of the type \textit{saddle/focus} and asymptotically press down upon the boundaries of the forbidden region. 

 \begin{figure}[h!]
 \centerline{\includegraphics[width=15cm,height=5cm]{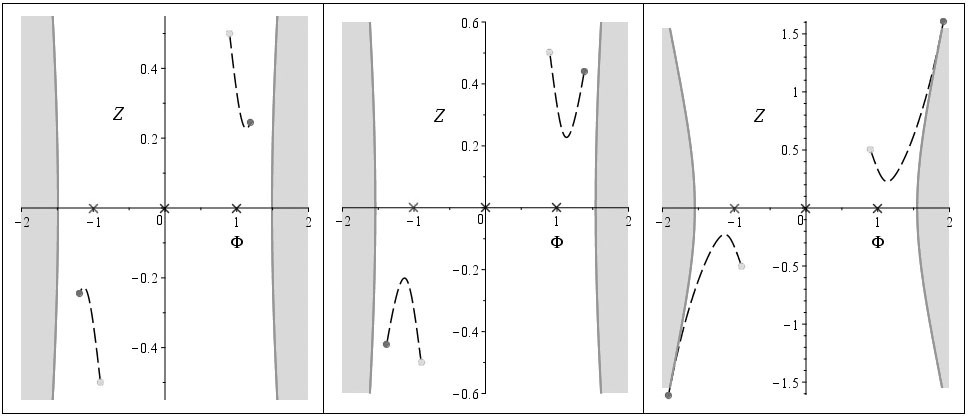}\label{Fig6}} \caption{Cosmological evolution of a scalar doublet with the parameters listed in Eq. \eqref{_4_} and the initial conditions listed in Eq. \eqref{_7_} in the \textit{classical} plane $\Sigma _{\Phi } \equiv \{ \Phi ,Z\} $. The phase diagrams correspond to the times (from left to right)  $\tau =1,\, \, 2\, \, {\rm and}\, \, 2.2489$.}
\end{figure} 

 \begin{figure}[h!]
 \centerline{\includegraphics[width=15cm,height=5cm]{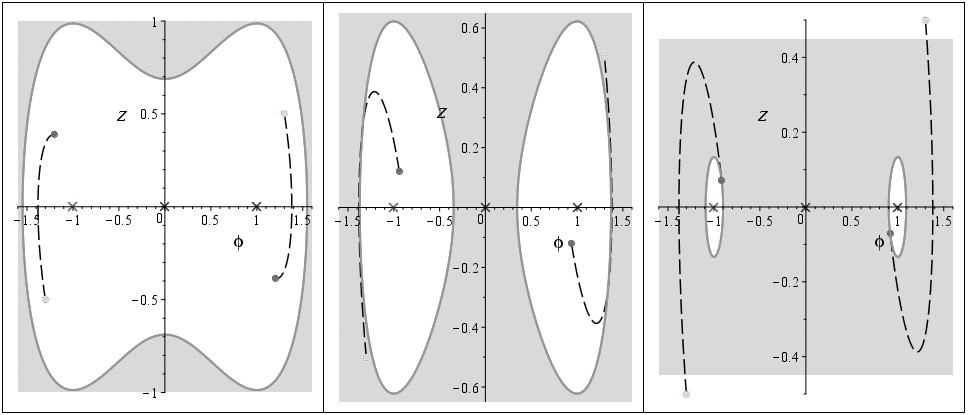}\label{Fig7}} \caption{Cosmological evolution of a scalar doublet with the parameters listed in Eq. \eqref{_4_} and the initial conditions listed in Eq. \eqref{_7_} in the \textit{phantom} plane $\Sigma _{\varphi } \equiv \{ \varphi ,z\} $. The phase diagrams correspond to the times (from left to right)  $\tau =1,\, \, 2\, \, {\rm and}\, \, 2.2489$.
}
\end{figure} 

 \begin{figure}[h!]
 \centerline{\includegraphics[width=15cm,height=5cm]{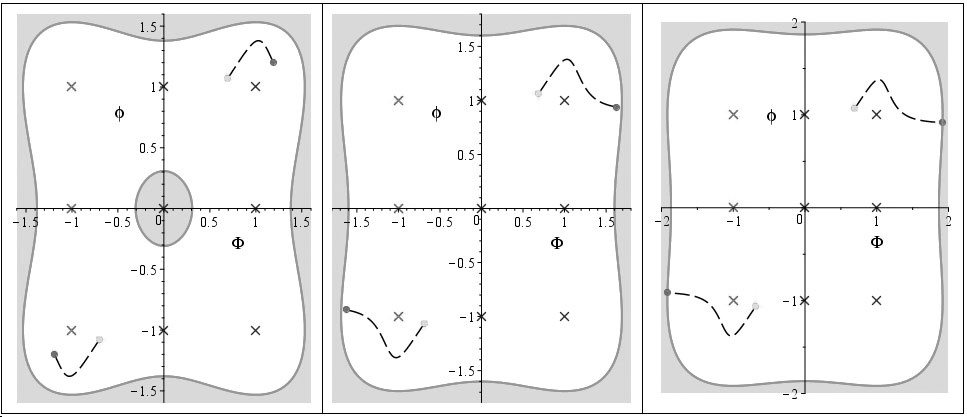}\label{Fig8}} \caption{Cosmological evolution of a scalar doublet with the parameters listed in Eq. \eqref{_4_} and the initial conditions listed in Eq. \eqref{_7_} in the plane of the potentials $\{ \Phi ,\varphi \} $.  The phase diagrams correspond to the times (from left to right) $\tau =1,\, \, 2\, \, {\rm and}\, \, 2.2489$.
}
\end{figure} 

Next, for simultaneous imaging of unlike-scale and unlike-sign quantities, we shall make use of a one-to-one, continuously differentiable function which we devised  ${\rm Lig}(x)$:
\[{\rm Lig}(x)\equiv {\rm sgn}(x)\log \, (1+|x|),\] 
which is such that     
\[{\rm Lig}(x)\approx \left\{\begin{array}{ll} {x,} & {|x|\to 0,} \\ {{\rm sgn}(x)\log |x|,} & {|x|\to \infty .} \end{array}\right. \] 
Figure 9  depicts the evolution of the physical characteristics of the cosmological model with the parameters listed in Eq. \eqref{_4_} and the initial conditions listed in Eq. \eqref{_7_}.  We see that the effective energy tends rapidly to zero, the pressure becomes positive, and the invariant cosmological acceleration tends to infinitely large negative values. Thus, there takes place a total and abrupt cessation of cosmological expansion, and the Universe becomes Euclidean.  The above examples show how different the behavior of the cosmological model can be, depending on its initial position relative to the singular points.  Note that the results shown in Fig. 9 describe the case of the transition of the inflationary Universe to a Euclidean one.  As we see, no Big Rip problem arises in our model.  

 \begin{figure}[h!]
 \centerline{\includegraphics[width=15cm,height=5cm]{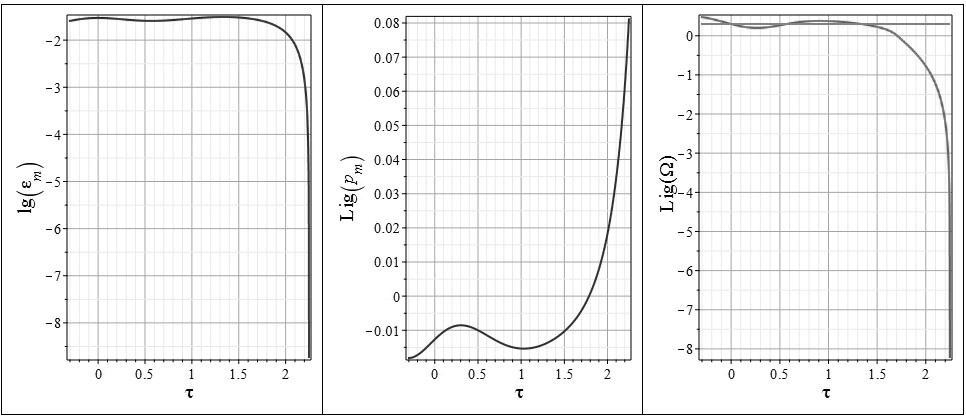}\label{Fig9}} \caption{Cosmological evolution of the physical characteristics of the cosmological model with the parameters listed in Eq. \eqref{_4_} and the initial conditions listed in Eq. \eqref{_7_}. From left to right:  the dimensionless effective energy ${\rm log}({\rm {\mathcal E}}_{m} )$, the dimensionless effective pressure ${\rm Lig}(p_{m} )$, and the invariant cosmological constant ${\rm Lig}(\Omega )$. The gray horizontal line in the middle of the graph corresponds to the value $\Omega =1$, i.e., to inflation. 
}
\end{figure}  

\subsubsection{Influence of the values of the parameters of the model   }

\noindent Let us elucidate how the absolute values of the parameters $\alpha _{m} $ and $\beta _{m} $, with the signs of all the parameters unchanged, influence the behavior of the model. As an example, let us consider the case  
\begin{equation} \label{_8_} 
{\bf P}=[10,10,1,1,1,0].   
\end{equation} 
The map of the singular points in this example coincides with the map in Fig. 1 [3] if we make the substitution $1\to 1/\sqrt{10} $ in the coordinates.  Figures 10--12 present results of a numerical modeling of dynamical system \eqref{_1_} for the parameters of the model listed in Eq. \eqref{_4_} and the initial conditions listed in Eq. \eqref{_9_} corresponding the initial values of the potentials: 
\begin{equation} \label{_9_} 
{\bf I}=[0.3,0,0.01,0].   
\end{equation} 
It is easy to see that this case does not differ qualitatively from the case considered above with the parameters listed in Eq. \eqref{_4_} (see Figs. 2--5).  The change in the initial conditions leads to analogous results.  The main factor here is accessibility of all the singular points. 

 \begin{figure}[h!]
 \centerline{\includegraphics[width=15cm,height=5cm]{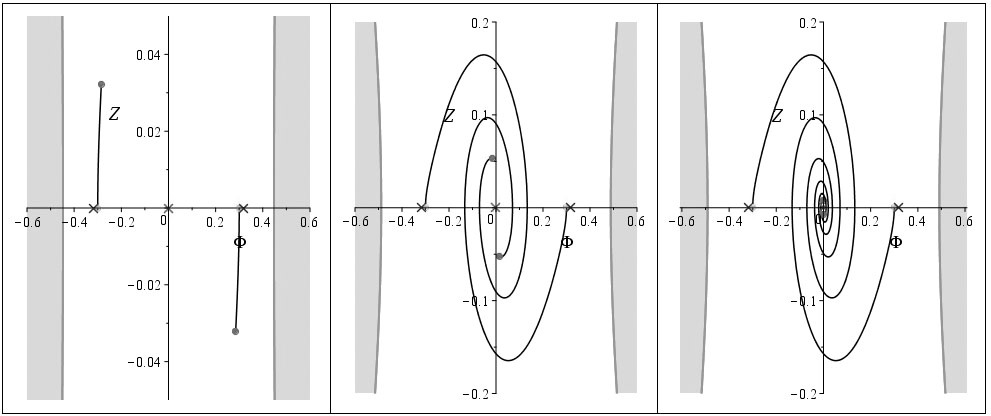}\label{Fig10}} \caption{Cosmological evolution of the scalar doublet with the parameters listed in Eq. \eqref{_4_} and the initial conditions listed in Eq. \eqref{_9_} in the \textit{classical} plane $\Sigma _{\Phi } \equiv \{ \Phi ,Z\} $. The phase diagrams correspond to the times (from left to right) $\tau =5,\, \, 10\, \, {\rm and}\, \, 20$. The forbidden regions correspond to the final time.
}
\end{figure}  

 \begin{figure}[h!]
 \centerline{\includegraphics[width=15cm,height=5cm]{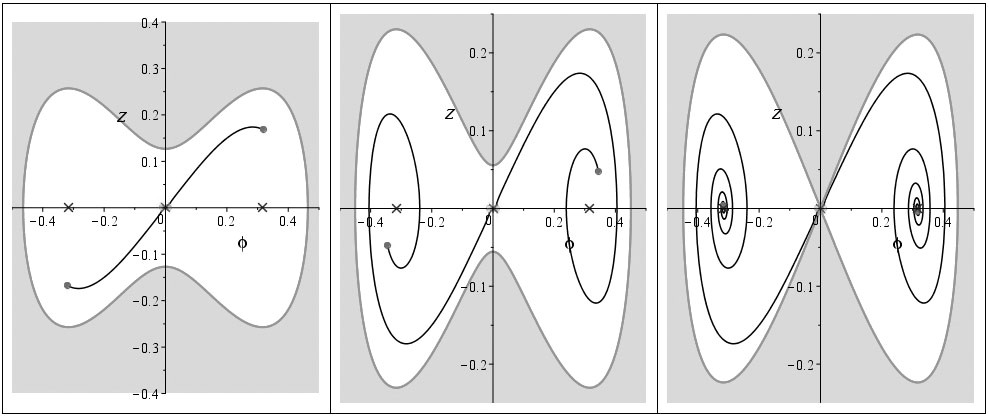}\label{Fig11}} \caption{Cosmological evolution of a scalar doublet with the parameters listed in Eq. \eqref{_4_} and the initial conditions listed in Eq. \eqref{_9_} in the phantom plane $\Sigma _{\varphi } \equiv \{ \varphi ,z\} $.
}
\end{figure} 
 \begin{figure}[h!]
 \centerline{\includegraphics[width=15cm,height=5cm]{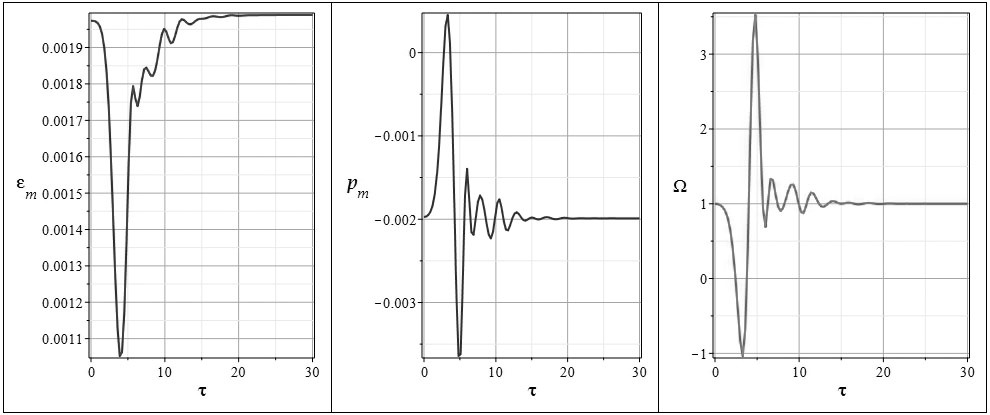}\label{Fig12}} \caption{ Cosmological evolution of the physical characteristics of the cosmological model with the parameters listed in Eq. \eqref{_4_} and the initial conditions listed in Eq. \eqref{_9_}.  From left to right: the dimensionless effective energy ${\mathcal E}_m$, the dimensionless effective pressure $p_{m} $, and the invariant cosmological acceleration $\Omega $.
}
\end{figure} 

\subsection{Bursts of cosmological acceleration    }

\noindent As was noted in [7--9], the presence of a phantom field in the cosmological model for small values of the potential of the phantom field leads to the appearance of phantom bursts of super-acceleration, characterized by large values of $\Omega $.  Figure 13 shows such bursts.  As we see, with decrease of the initial value of the phantom field potential, the burst of acceleration takes place at a later time with simultaneous growth of its amplitude. After the burst, the model enters the inflation stage.  
 \begin{figure}[h!]
 \centerline{\includegraphics[width=8.5cm]{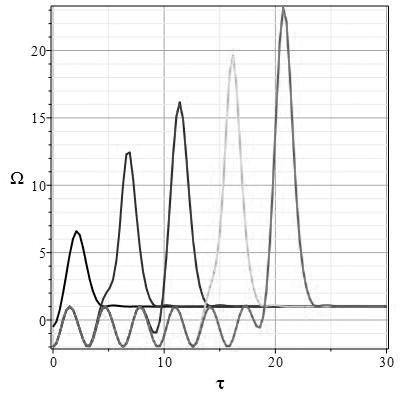}\label{Fig13}} \caption{  Bursts of cosmological acceleration $\Omega $ for the parameters of the model listed in Eq. \eqref{_4_} for the initial value of the potential of the classical field and its derivative  $\Phi (0)=0.1,\, \, Z(0)=0,\, \, z(0)=0$. From left to right:  $\varphi (0)=10^{-1} $, $\varphi (0)=10^{-3} $, $\varphi (0)=10^{-5} $, $\varphi (0)=10^{-7} $ and $\varphi (0)=10^{-9} $.
}
\end{figure}

\section{Analysis of the results   }

\noindent As our investigations have shown, dynamical system \eqref{_1_} can have three types of behavior: 

\noindent {\bf I. Standard}.  The phase trajectories in both planes, $\Sigma _{\Phi } $ and $\Sigma _{\varphi } $, spiral around and onto the centers/foci.  This case can be broken down into three subcases: 

1. The phase trajectories in both planes wind around and onto the accessible zero center $M_{0} $.  This case corresponds to the parameters listed in Eq. \eqref{_10_} and the initial conditions listed in Eq. \eqref{_11_}.  Here the pressure of the dynamical system tends to zero and the cosmological acceleration executes oscillations with amplitude on the order of 1 near the value $\Omega _{0} =-1/2$, corresponding to the nonrelativistic equation of state.  

2. The phase trajectories in the \textit{classical} plane wind around and onto the accessible nonzero focus $M_{0} $, whereas the phase trajectories in the phantom plane wind around and onto the accessible nonzero foci $M_{01} \; {\rm and}\; M_{02} $.  This case corresponds, for example, to the parameters listed in Eq. \eqref{_4_} and the initial conditions listed in Eq. \eqref{_6_} (see Figs. 3--6);  to the parameters listed in Eq. \eqref{_4_} and the initial conditions listed in Eq. \eqref{_9_} (see Figs. 11--13).  In all of these cases, the magnitude of the cosmological acceleration has a jump $\Omega _{{\rm max}} >1$, after which the cosmological model enters the inflationary regime $\Omega =1$. 

3. The phase trajectories in the \textit{classical} plane wind around and onto the inaccessible nonzero centers/foci $M_{10} \; {\rm and}\; M_{20} $, whereas the phase trajectories in the phantom plane wind around and onto the inaccessible nonzero centers $M_{01} \; {\rm and}\; M_{02} $, and in both planes the phase trajectories attempt to adhere to the boundary of the region with zero effective energy, but in the final count they do not succeed in doing so.  As a result, the cosmological acceleration executes anharmonic oscillations with large amplitude around the value $\Omega _{0} =-1$.  This case corresponds to the parameters listed in Eq. \eqref{_14_} and the initial conditions listed in Eq. \eqref{_15_} (see Figs. 23--25).

\noindent {\bf II. Rebound}. The phase trajectories in the $\Sigma _{\Phi } $ plane tend asymptotically to the line $Z=0$, whereas the phase trajectories in the $\Sigma _{\varphi } $ plane are repulsed from the saddle points $M_{01} \; {\rm and}\; M_{02} $ and escape to infinity along the asymptote $\varphi =z,\tau \to \infty $. The cosmological acceleration after the burst enters the inflation regime -- the model parameters listed in Eq. \eqref{_10_} and the initial conditions listed in Eq. \eqref{_13_} (see Figs. 20--22).

\noindent {\bf III. Adherence}.  The phase trajectories in both planes, $\Sigma _{\Phi } $ and $\Sigma _{\varphi } $, press down upon the zero effective energy hypersurfaces.   This happens because of repulsion of the phase trajectory from a saddle point (See Fig. 7, for example) or because of attraction to an inaccessible center/focus. The effective energy falls rapidly to zero and the cosmological acceleration tends to $-\infty $ -- a very rapid braking takes place: the parameters listed in Eq. \eqref{_4_} and the initial conditions listed in Eq. \eqref{_7_} (Figs.~7--10), the parameters listed in Eq. \eqref{_10_} and the initial conditions listed in Eq. \eqref{_12_} (see Figs. 17--19).

\section{Some results of numerical modeling  }

\subsection{Case of inaccessibility of the singular points $M_{01} $ and $M_{02} $}
\begin{equation} \label{_10_} 
{\bf P}=[1,-1,1,-1,1,0].   
\end{equation} 

\subsubsection{General properties of the phase space    }

\noindent In this case, the singular points have the same coordinates as in the previous case \eqref{_5_}, but the invariant characteristics $\sigma _{i}^{2} $ are equal to    
\[\sigma _{1}^{2} =-\frac{3}{8} ,\quad \sigma _{2}^{2} =\frac{3}{8} ,\quad \sigma _{3}^{2} =0.\] 
The singular points $M_{01} $ and $M_{02} $ of the dynamical system are inaccessible.   

\subsubsection{Phase trajectories of the dynamical system     }

\noindent Figures 14--16 present results of numerical modeling of dynamical system \eqref{_1_} for the model parameters listed in Eq. \eqref{_10_} and the initial conditions listed in Eq. \eqref{_11_}: 
\begin{equation} \label{_11_} 
{\bf I}=[0.7,0.5,0.1,0].   
\end{equation} 
 \begin{figure}[h!]
 \centerline{\includegraphics[width=15cm,height=5cm]{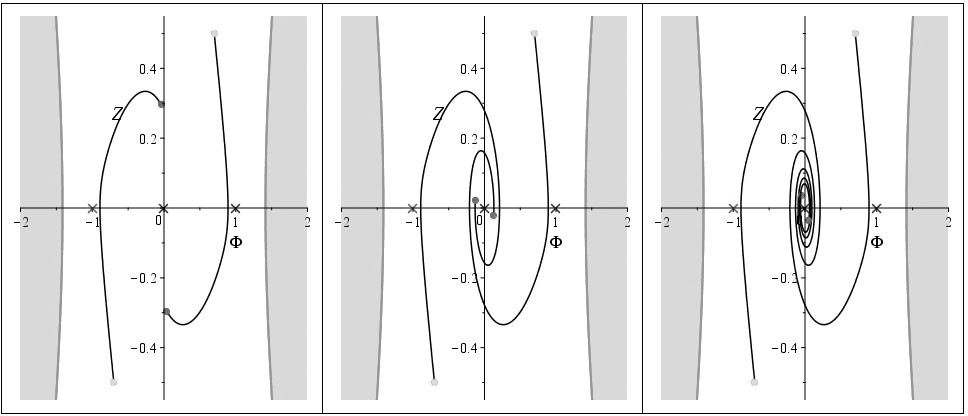}\label{Fig14}} \caption{ Cosmological evolution of a scalar doublet with the parameters listed in Eq. \eqref{_10_} and the initial conditions listed in Eq. \eqref{_11_} in the \textit{classical} plane $\Sigma _{\Phi } \equiv \{ \Phi ,Z\} $. The phase diagrams correspond to the times (from left to right)  $\tau =5,\, \, 10\, \, {\rm and}\, \, 20$.
}
\end{figure}

 \begin{figure}[h!]
 \centerline{\includegraphics[width=15cm,height=4.5cm]{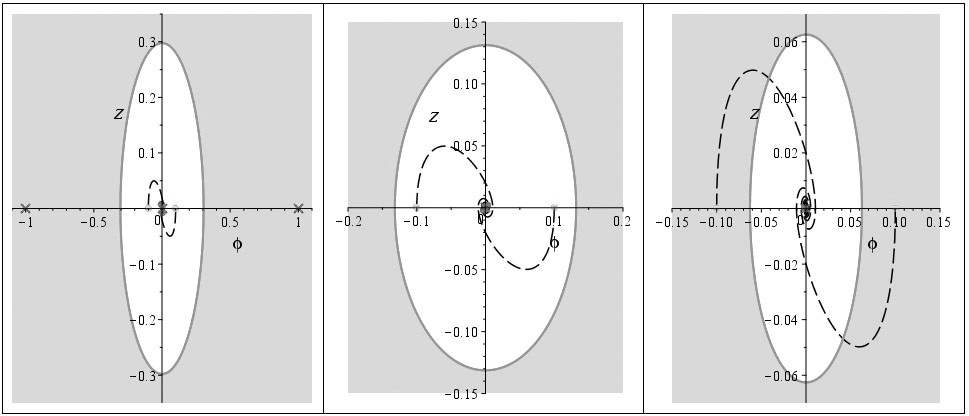}\label{Fig15}} \caption{Cosmological evolution of a scalar doublet with the parameters listed in Eq. \eqref{_10_} and the initial conditions listed in Eq. \eqref{_11_} in the phantom plane $\Sigma _{\varphi } \equiv \{ \varphi ,z\} $.  The phase diagrams correspond to the times (from left to right)    $\tau =5,\, \, 10\, \, {\rm and}\, \, 20$.
}
\end{figure}

 \begin{figure}[h!]
 \centerline{\includegraphics[width=15cm,height=5cm]{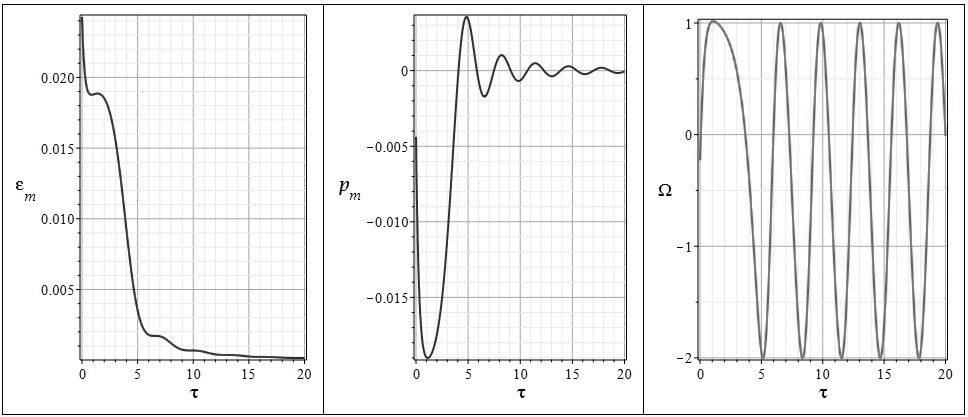}\label{Fig16}} \caption{ Cosmological evolution of the physical characteristics of the cosmological model with the parameters listed in Eq. \eqref{_10_} and the initial conditions listed in Eq. \eqref{_11_}. From left to right:  the dimensionless effective energy ${\rm log}({\rm {\mathcal E}})$, the dimensionless effective pressure ${\rm Lig}(p_{m} )$, and the invariant cosmological acceleration ${\rm Lig}(\Omega )$. The gray horizontal line in the last graph corresponds to the value $\Omega =1$, i.e., to inflation.
}
\end{figure}

\noindent Thus, for the initial conditions listed in Eq. \eqref{_11_}, the phase trajectories of the dynamical system in both planes, $\Sigma _{\Phi } $ and $\Sigma _{\varphi } $, wind around and onto the zero center  $M_{0} $.  Here the effective energy and pressure of the system tend to zero while the invariant cosmological acceleration executes oscillations around the value $\Omega =-1/2$ corresponding to the macroscopic nonrelativistic equation of state ${\rm \kappa }=0$.  Precisely speaking, the same oscillations arise in the model with a classical scalar field (see [10, 11]). Figures 17--19 present results of numerical modeling of dynamical system \eqref{_1_} for the model parameters listed in Eq. \eqref{_10_} and the initial conditions listed in Eq. \eqref{_12_}:  
\begin{equation} \label{_12_} 
{\bf I}=[0.7,0.5,0.9,0].   
\end{equation}

 \begin{figure}[h!]
 \centerline{\includegraphics[width=15cm,height=5cm]{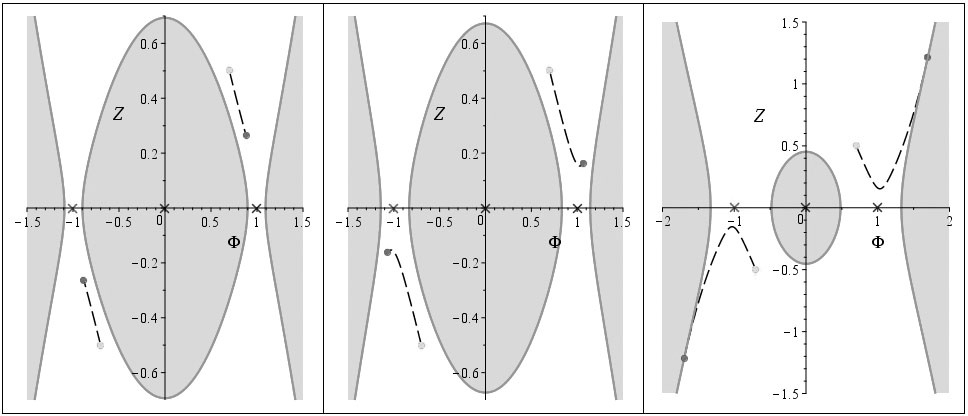}\label{Fig17}} \caption{Cosmological evolution of a scalar doublet with the parameters listed in Eq. \eqref{_10_} and the initial conditions listed in Eq. \eqref{_12_} in the \textit{classical} plane $\Sigma _{\Phi } \equiv \{ \Phi ,Z\} $. The phase diagrams correspond to the times (from left to right)     $\tau =0.5,\, \, 1.5\, \, {\rm and}\, \, 2.99$.
}
\end{figure}
 \begin{figure}[h!]
 \centerline{\includegraphics[width=15cm,height=5cm]{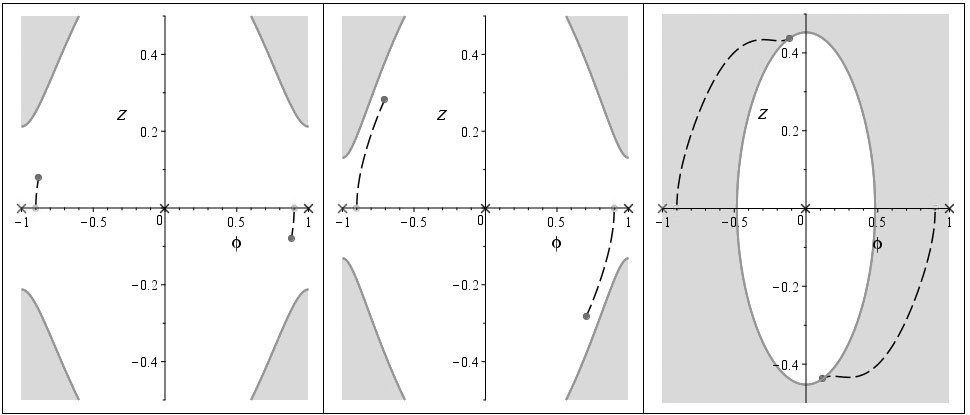}\label{Fig18}} \caption{Cosmological evolution of a scalar doublet with the parameters listed in Eq. \eqref{_10_} and the initial conditions listed in Eq. \eqref{_12_} in the phantom plane $\Sigma _{\varphi } \equiv \{ \varphi ,z\} $.  The phase diagrams correspond to the times (from left to right)     $\tau =0.5,\, \, 1.5\, \, {\rm and}\, \, 2.99$.
}
\end{figure}

\noindent Figures 20 and 21 present results of numerical modeling of dynamical system \eqref{_1_} for the model parameters listed in Eq. \eqref{_10_} and the initial conditions listed in Eq. \eqref{_13_}:    
\begin{equation} \label{_13_} 
{\bf I}=[0.7,0.5,1.5,0],   
\end{equation} 
when the dynamical system starts from a position higher than the singular point $M_{01} $.  This case can be called a \textit{rebound} -- in the phantom plane the system rebounds from the forbidden region and rapidly grows the potential and kinetic energy of the phantom field, in the process entering the inflationary regime of acceleration.  

 \begin{figure}[h!]
 \centerline{\includegraphics[width=15cm,height=5cm]{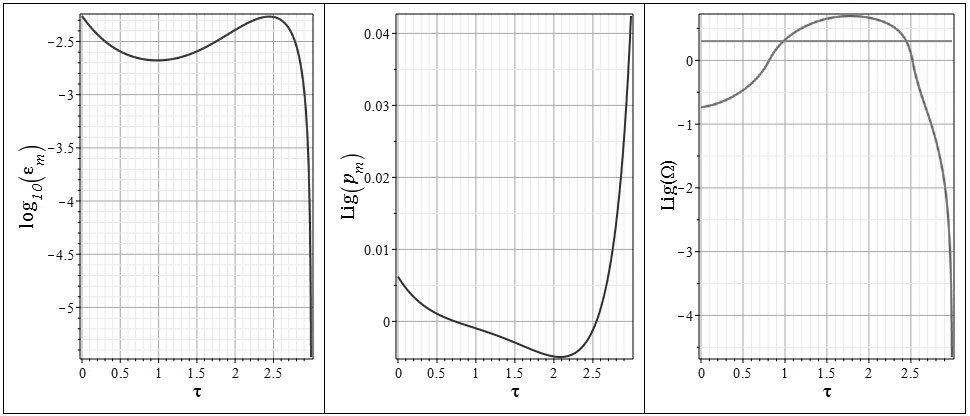}\label{Fig19}} \caption{ Cosmological evolution of the physical characteristics of the cosmological model  with the parameters listed in Eq. \eqref{_10_} and the initial conditions listed in Eq. \eqref{_12_}.  From left to right:   the dimensionless effective energy ${\rm log}({\rm {\mathcal E}})$, the dimensionless effective pressure ${\rm Lig}(p_{m} )$, and the invariant cosmological acceleration ${\rm Lig}(\Omega )$. The gray horizontal line in the last graph corresponds to the value $\Omega =1$, i.e., to inflation.
}
\end{figure}
\noindent 

\begin{figure}[h!]
 \centerline{\includegraphics[width=15cm,height=5cm]{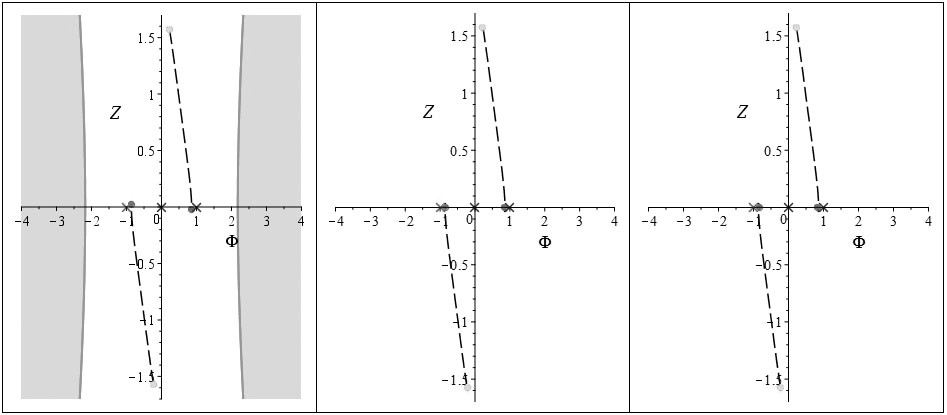}\label{Fig20}} \caption{Cosmological evolution of a scalar doublet with the parameters listed in Eq. \eqref{_10_} and the initial conditions listed in Eq. \eqref{_13_} in the \textit{classical} plane $\Sigma _{\Phi } \equiv \{ \Phi ,Z\} $.  The phase diagrams corresponds to the times (from left to right)     $\tau ={\rm 1,}\, \, {\rm 5}\, \, {\rm and}\, \, {\rm 8.79}$.
}
\end{figure}

\begin{figure}[h!]
 \centerline{\includegraphics[width=15cm,height=5cm]{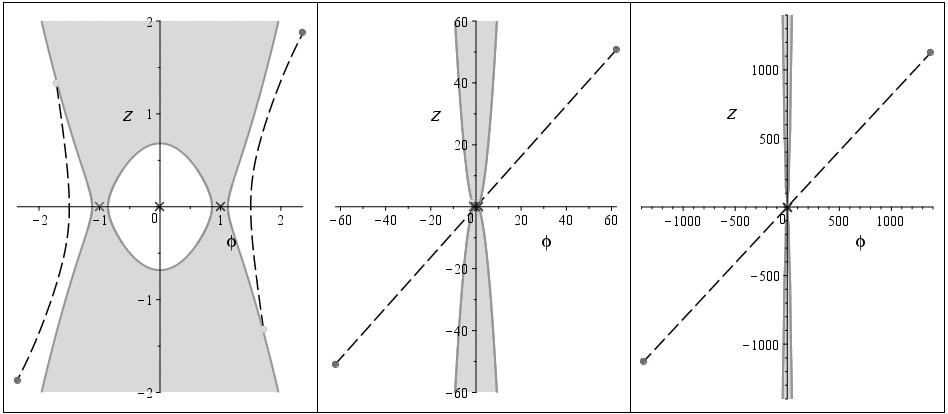}\label{Fig21}} \caption{Cosmological evolution of a scalar doublet with the parameters listed in Eq. \eqref{_10_} and the initial conditions listed in Eq. \eqref{_13_} in the phantom plane $\Sigma _{\varphi } \equiv \{ \varphi ,z\} $.  The phase diagrams corresponds to the times (from left to right)  $\tau ={\rm 1,}\, \, {\rm 5}\, \, {\rm and}\, \, {\rm 8.79}$.
}
\end{figure}

\begin{figure}[h!]
 \centerline{\includegraphics[width=15cm,height=5cm]{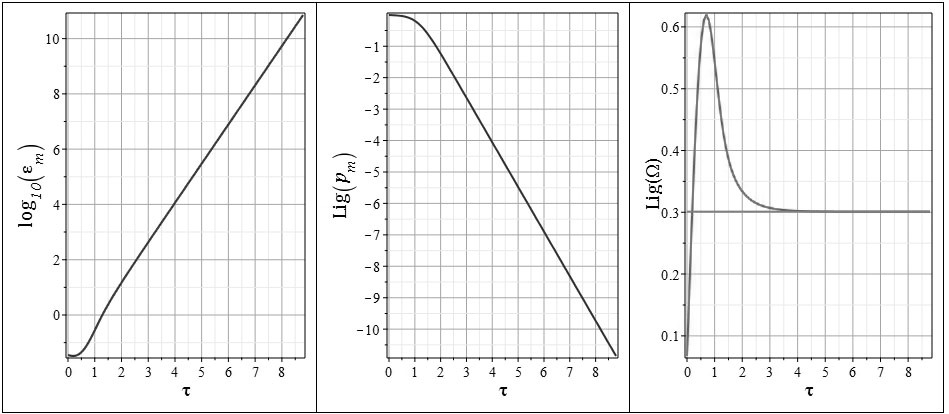}\label{Fig22}} \caption{Cosmological evolution of the physical characteristics of the cosmological model  with the parameters listed in Eq. \eqref{_10_} and the initial conditions listed in Eq. \eqref{_13_}. From left to right: the dimensionless effective energy ${\rm log}({\rm {\mathcal E}})$, the dimensionless effective pressure ${\rm Lig}(p_{m} )$, and the invariant cosmological acceleration ${\rm Lig}(\Omega )$.  The gray horizontal line in the last graph corresponds to the value $\Omega =1$, i.e., to inflation.
}
\end{figure}

\noindent Figure 22 depicts the evolution of the physical characteristics of the cosmological model with the parameters listed in Eq. \eqref{_10_} and the initial conditions listed in Eq. \eqref{_13_}.

\noindent 
\subsection {Case of inaccessibility of the singular points $M_{10} $ and $M_{20} $}
\begin{equation} \label{_14_} 
{\bf P}=[-1,1,-1,1,1,0].   
\end{equation} 

\subsubsection{ General properties of the phase space    }

\noindent In this case, the singular points have the same coordinates as in the previous case \eqref{_5_}, and the invariant characteristics $\sigma _{i}^{2} $ are equal to    
\[\sigma _{1}^{2} =\frac{3}{8} ,\quad \sigma _{2}^{2} =-\frac{3}{8} ,\quad \sigma _{3}^{2} =0.\] 
The singular points $M_{10} $ and $M_{20} $ of the dynamical system are inaccessible.  

\noindent 
\subsubsection{ Phase trajectories of the dynamical system      }

\noindent Figures 23--25 present results of numerical modeling of dynamical system \eqref{_1_} for the model parameters listed in Eq. \eqref{_14_} and the initial conditions listed in Eq. \eqref{_15_}: 
\begin{equation} \label{_15_} 
{\bf I}=[0.5,0.5,0.5,0].   
\end{equation} 
\begin{figure}[h!]
 \centerline{\includegraphics[width=15cm,height=5cm]{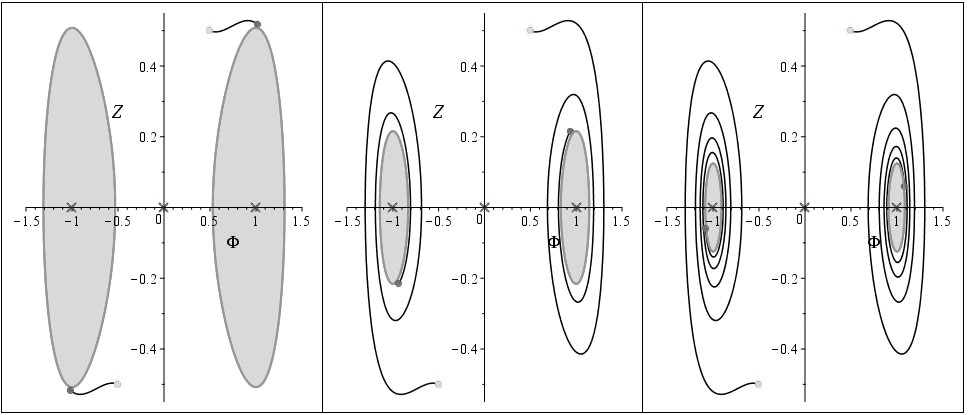}\label{Fig23}} \caption{Cosmological evolution of a scalar doublet with the parameters listed in Eq. \eqref{_14_} and the initial conditions listed in Eq. \eqref{_15_} in the \textit{classical} plane $\Sigma _{\Phi } \equiv \{ \Phi ,Z\} $.  The phase diagrams correspond to the times (from left to right)    $\tau =5,\, \, 10\, \, {\rm and}\, \, 20$.
}
\end{figure}
\begin{figure}[h!]
 \centerline{\includegraphics[width=15cm,height=5cm]{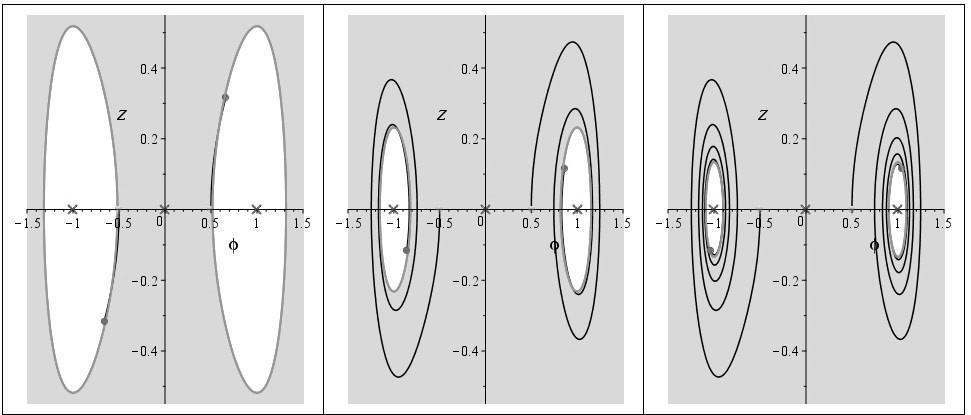}\label{Fig24}} \caption{Cosmological evolution of a scalar doublet with the parameters listed in Eq. \eqref{_14_} and the initial conditions listed in Eq. \eqref{_15_} in the \textit{phantom} plane $\Sigma _{\varphi } \equiv \{ \varphi ,z\} $.  The phase diagrams correspond to the times (from left to right)   $\tau =5,\, \, \, 10\, \, {\rm and}\, \, 20$.
}
\end{figure}
\begin{figure}[h!]
 \centerline{\includegraphics[width=15cm,height=5cm,height=5cm]{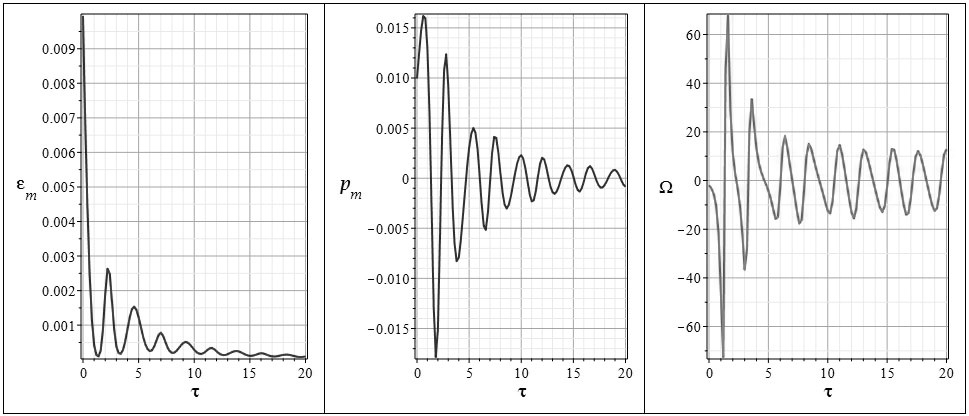}\label{Fig25}} \caption{Cosmological evolution of the physical characteristics of the cosmological model with the parameters listed in Eq. \eqref{_14_} and the initial conditions listed in Eq. \eqref{_15_}.
}
\end{figure}

\noindent It can be said that the behavior of the dynamical system has practically no dependence on the initial conditions despite the fact that the geography of the forbidden regions can be substantially different, which is probable evidence of stability of the behavior of the model for the parameters listed in Eq. \eqref{_14_} with respect to a change in the initial conditions.  A study of the behavior of the dynamical system under an increase by a factor of 10 of the absolute values of the model parameters $\alpha _{m} \; {\rm and}\; \, \beta _{m} $, but with their signs left unchanged ${\bf P}=[10,-10,1,-1,1,0]$, leads to qualitatively the same results with the one difference being that the amplitude of the oscillations of the cosmological acceleration decreases abruptly and is on the order of 1  (in Fig. 25 it exceeds 60).   

\noindent 
\section{Conclusions   }

\noindent Summarizing the results of the study presented here, we note that the peculiarities of dynamical system \eqref{_1_} revealed in this work are associated, first of all, with attraction of its phase trajectories to zero effective energy hypersurfaces and can form the basis of new cosmological scenarios. Indeed, the passage of the cosmological model to stationary orbits with zero effective energy for nonzero potentials of the scalar fields of an asymmetric doublet and their first derivatives allows us to treat these orbits as purely vacuum states of these fields, corresponding to zero curvature of Friedmann spacetime, i.e., to a purely Euclidean space.  This space, as it turns out, can be non-empty and contain matter in the form of a pair of synchronously oscillating scalar fields, which can be considered as virtual vacuum fields.   Note that all of the considered cases of adherence of the dynamical system to the zero effective energy surfaces correspond to very early stages of cosmological evolution.  The possible instability of these scalar fields, stationary with respect to perturbations, can become a source of creation of real particles as a result of scalar interactions, not the result of a gravitational instability.

This work was performed within the scope of the Russian Government Program of Competitive Growth of Kazan Federal University.


\begin{thebibliography}{15}
\bibitem{1}  Yu.~G. Ignat'ev and I.~A. Kokh, Russ. Phys. J., \textbf{61}, No. 6, 1079--1092 (2018).
\bibitem{2} Yu.~G. Ignat'ev and I.~A. Kokh, Russ. Phys. J., \textbf{61}, No. 9,  1590--1596 (2018).
\bibitem{3} Yu.~G. Ignat'ev and I.~A. Kokh, Russ. Phys. J., \textbf{62}, No. 2,  in press (2019). 
\bibitem{4} I.Ya. Aref'eva, N.~V. Bulatov, R.~V. Gorbachev, and S. Yu. Vernov, Class. Quantum Grav. -- 2014. -- V. \textbf{31}, 065007. 
\bibitem{5} Yu. Ignat'ev, A. Agathonov, M. Mikhailov, and D. Ignatyev, Astr. Space Sci., \textbf{357}, 61 (2015).
\bibitem{6} Yurii Ignat'ev, Alexander Agathonov, and Dmitry Ignatyev, arXiv:1608.05020 [gr-qc] (2016).
\bibitem{7} Yu.~G. Ignat'ev, A.~A. Agathonov, and D. Yu. Ignatyev, Grav. Cosmol. -- 2018. -- V. \textbf{24},~1, 1--12.
\bibitem{8} Yurii Ignat'ev, Alexander Agathonov, and Irina Kokh, arXiv:1810.09873 [gr-qc].
\bibitem{9} Yu.~G. Ignat'ev and A. R. Samigullina, Space, Time and Fund. Interact., No. 1, 100--102 (2017). 
\bibitem{10} Yu.~G. Ignat'ev and A. R. Samigullina, Russ. Phys. J., \textbf{60}, 7,  1173--1181 (2017).
\bibitem{11} Yu.~G. Ignat'ev, D. Yu. Ignatyev, and A.~R. Samigullina, Grav. Cosmol., \textbf{24}, 2, 148--153 (2018); arXiv:1705.05000 [gr-qc].

\end{thebibliography}
\end{document}